\documentstyle[prb,aps]{revtex}
\begin{document}

\title{Computing Number Fluctuations}
\author{Girish S. Setlur and Yia-Chung Chang}
\address{Department of Physics and Materials Research Laboratory,\\
 University of Illinois at Urbana-Champaign , Urbana Il 61801}
\maketitle

\begin{abstract}
 Here we try and delienate the properties of the function that corresponds
 to fluctuations in the momentum distribution. The quantity denoted 
 by $ N(k,k^{'}) $ is quite an interesting object.
 It satisfies various elegant
 sum rules and is also quite singular in some respects. All these properties
 are brought out and a formal connection is found between this object
 and the momentum distribution of the interacting and non-interacting
 many-fermion systems. This exercise is quite general in that it does
 not refer to any particular hamiltonian. It is also quite useful since
 in an earlier preprint(cond-mat/9810043) we showed how to compute
 the spectral function and single-particle lifetime of homogeneous
 Fermi systems where the only undetermined quantity was this function
 $ N(k,k^{'}) $.
\end{abstract}

\section{Number Fluctuations}

 Without further ado, let us define the object of interest, and try and
 enumerate its many interesting properties.
 The number-fluctuation $ N({\bf{k}},{\bf{k}}^{'}) $ is defined as,
\begin{equation}
N({\bf{k}},{\bf{k}}^{'}) = \langle n_{ {\bf{k}} }n_{ {\bf{k}}^{'} } \rangle
 - \langle n_{ {\bf{k}} } \rangle  \langle n_{ {\bf{k}}^{'} } \rangle
\end{equation}
 here $ n_{ {\bf{k}} } = c^{\dagger}_{ {\bf{k}} }c_{ {\bf{k}} } $.
 $ c_{ {\bf{k}} } $ and $ c^{\dagger}_{ {\bf{k}} } $ are Fermi fields.
 It is useful at the outset to point out some rather trivial
 facts about this object,
\begin{equation}
\sum_{ {\bf{k}} }N({\bf{k}},{\bf{k}}^{'}) = 0
\end{equation}
 This has to do with the fact that the total number of particles is conserved.
 But for those who are skeptical a simple demonstration may settle any
 lingering doubts. Also this approach will be used for illustrating more
 substantive issues later on, hence its introduction at this stage is quite
 appropriate.
\begin{equation}
\sum_{ {\bf{k}} }N({\bf{k}},{\bf{k}}^{'}) = 
\langle N\mbox{         }n_{ {\bf{k}}^{'} } \rangle
 - \langle N\rangle\langle n_{ {\bf{k}}^{'} } \rangle
\end{equation}
 Let $ \{ | i \rangle_{0} \} $ be the eigenstates of
 the noninteracting Fermi system
 with total number of particles equal to $ N_{0} $. And
 $ \{ | i \rangle \} $  be the eigenstates of the interacting Fermi system
 with total number of particles equal to $ N_{0} $.
 We may write,
\begin{equation}
\langle N\mbox{         }n_{ {\bf{k}}^{'} } \rangle
 = \frac{1}{Z}
\sum_{j} e^{-\beta E_{j}}\langle j |N\mbox{         }n_{ {\bf{k}}^{'} } 
| j \rangle
\end{equation}
here $ Z = \sum_{j} e^{-\beta E_{j}} $ is the canonical partition function.
 But,
\begin{equation}
\langle j |N\mbox{         }n_{ {\bf{k}}^{'} }| j \rangle
 = \sum_{i} \langle j |N\mbox{         }| i \rangle
 \langle i |n_{ {\bf{k}}^{'} }| j \rangle
 = N_{0} \langle j |n_{ {\bf{k}}^{'} }| j \rangle
\end{equation}
 Since, $ \langle N \rangle = N_{0} $ also, the result follows. 
 A similar result holds for the "first moment" namely,
 $ \sum_{ {\bf{k}} }{\bf{k}}N({\bf{k}},{\bf{k}}^{'}) $.
 Since $ {\bf{P}} = \sum_{ {\bf{k}} }{\bf{k}}n_{ {\bf{k}} } $
 is the total momentum of electrons, the same argument gives us,
 $ \sum_{ {\bf{k}} }{\bf{k}} N({\bf{k}},{\bf{k}}^{'}) = 0 $.
 But what about those cases when the eigenstates of the interacting system
 are not eigenstates of the operator $ {\bf{P}} $ ? This case for example
 corresponds to the situation when translational invariance is broken
 spontaneously(Wigner crystal). The answer is as follows. 
\begin{equation}
\langle j | {\bf{P}}\mbox{         }n_{ {\bf{k}}^{'} }| j \rangle
 = \sum_{i_{0}} \langle j |{\bf{P}}\mbox{         }| i_{0} \rangle
 \langle i_{0} |n_{ {\bf{k}}^{'} }| j \rangle
 = \sum_{i_{0}} n^{i0}_{ {\bf{k}}^{'} }{\bf{P}}_{i0}
|\langle i_{0} | j \rangle|^{2}
\end{equation}
 here $ \{ | i0 \rangle \} $ are the states of the free theory.
 Therefore $ {\bf{P}}_{i0} = 0 $ for all $ i0 $. But then
 the overlap $ \langle i_{0} | j \rangle $ may be singular.
 By that I mean it may be actually infinite over a set of measure
 zero. Thereby orthonormality is not compromised and yet
 $ \sum_{ {\bf{k}} }{\bf{k}}N({\bf{k}},{\bf{k}}^{'}) $ may fail to
 vanish.
 Let us prove another interesting property of this function.
\begin{equation}
N({\bf{k}},{\bf{k}}) = \langle n^{2}_{ {\bf{k}} }  \rangle 
 - \langle n_{ {\bf{k}} }  \rangle^{2}
 =  \langle n_{ {\bf{k}} }  \rangle( 1 - \langle n_{ {\bf{k}} }  \rangle)
\end{equation}  
 A quick proof of this involves making the following observation,
 $ n^{2}_{ {\bf{k}} } = n_{ {\bf{k}} } $. But this equation has
 to be interpreted with care. It is not an operator identity in the
 usual sense. If it were, many strange (and incorrect) results would ensue.
 For example( $ |{\bf{q}}| << |{\bf{k}}| $), 
\[
\langle n_{ {\bf{k+q}} } \rangle \approx
\langle n_{ {\bf{k}} } \rangle 
 +\langle ({\bf{q}}.\nabla_{ {\bf{k}} }) n_{ {\bf{k}} } \rangle
\]
\[
 = \langle n_{ {\bf{k}} } \rangle
 +\langle  ({\bf{q}}.\nabla_{ {\bf{k}} }) n^{2}_{ {\bf{k}} } \rangle
 =  \langle n_{ {\bf{k}} } \rangle
 + 2\langle  n_{ {\bf{k}} }({\bf{q}}.\nabla_{ {\bf{k}} }) n_{ {\bf{k}} } \rangle
\]
(because $ ({\bf{q}}.\nabla_{ {\bf{k}} }) n_{ {\bf{k}} }  \approx  n_{ {\bf{k+q}} } -  n_{ {\bf{k}} } $
and $ [n_{ {\bf{k}} }, ({\bf{q}}.\nabla_{ {\bf{k}} }) n_{ {\bf{k}} }] = 0 $)
\[
 =  \langle n_{ {\bf{k}} } \rangle
+ 2\langle  n_{ {\bf{k}} }( n_{ {\bf{k+q}} }- n_{ {\bf{k}} }) \rangle
= 2\langle  n_{ {\bf{k}} }n_{ {\bf{k+q}} }\rangle
-\langle n_{ {\bf{k}} } \rangle
\]
This means,
\[
  \langle  n_{ {\bf{k}} }n_{ {\bf{k+q}} }\rangle 
 -  \langle  n_{ {\bf{k}} } \rangle \langle n_{ {\bf{k+q}} }\rangle
 = 
-\frac{1}{2}(-\langle n_{ {\bf{k+q}} } \rangle 
- \langle n_{ {\bf{k}} } \rangle
 +  2\langle  n_{ {\bf{k}} } \rangle \langle n_{ {\bf{k+q}} }\rangle)
\]
 We can see quite easily that this equation is false by applying it to
 the noninteracting system at zero temperature.
 The left side is identically zero and right side may not be.
 However, we may still apply the ideas introduced earlier to write down
 some useful formulas for these objects(N's). Then in what sense is
 $ n^{2}_{ {\bf{k}} } = n_{ {\bf{k}} } $ true ? The answer is, in the
 sense of matrix elements.
\begin{equation}
\langle i_{0}|n^{2}_{ {\bf{k}} } | j_{0} \rangle
 = \sum_{k_{0}}\langle i_{0}|n_{ {\bf{k}} }
| k_{0} \rangle
 \langle k_{0}|n_{ {\bf{k}} } | j_{0} \rangle
 = n^{i0}_{ {\bf{k}} }n^{j0}_{ {\bf{k}} }
\sum_{k_{0}}\delta_{ i_{0}, k_{0} }
\delta_{ j_{0}, k_{0} }
 = (n^{i0}_{ {\bf{k}} })^{2}\delta_{ i_{0}, j_{0} }
\end{equation}
 Since the c-number $ n^{i0}_{ {\bf{k}} } $ is either zero or one,
 the result follows.
 A further subtle issue comes
 to mind. In the case of noninteracting systems at finite temperature,
 $ N({\bf{k}},{\bf{k}}) \neq 0 $ but $  N({\bf{k}},{\bf{k}}^{'}) = 0 $ 
 for $ {\bf{k}} \neq {\bf{k}}^{'} $. Therefore $ N({\bf{k}},{\bf{k}}^{'}) $
 is discontinuous at $ {\bf{k}} = {\bf{k}}^{'} $. In particular,
\begin{equation}
Lim_{ |{\bf{q}}|, |{\bf{q}}|^{'}  \rightarrow 0}
 N({\bf{k-q}},{\bf{k}}-{\bf{q}}^{'}) \neq
N({\bf{k}},{\bf{k}}) 
\end{equation}
 Having done all this we now proceed to write down formulas for the
 off-diagonal elements of $ N $.
\begin{equation}
N({\bf{k}},{\bf{k}}^{'}) = \frac{1}{Z}
\sum_{i}e^{-\beta E_{i}}\langle i |n_{ {\bf{k}} }n_{ {\bf{k}}^{'} }
| i \rangle
-\langle n_{ {\bf{k}} } \rangle \langle n_{ {\bf{k}}^{'} } \rangle
\end{equation}
here $ | i \rangle $ are the eigenstates of the full theory. We may now
 decompose 
\begin{equation}
\langle i |n_{ {\bf{k}} }n_{ {\bf{k}}^{'} }
| i \rangle = 
\sum_{j_{0}}\langle i |n_{ {\bf{k}} } |j_{0} \rangle
\langle j_{0}| n_{ {\bf{k}}^{'} }| i \rangle
 = \sum_{j_{0}}n^{j0}_{ {\bf{k}} }n^{j0}_{ {\bf{k}}^{'} }
|\langle j_{0}|i \rangle|^{2}
\end{equation}
Further,
\begin{equation}
\langle n_{ {\bf{k}} } \rangle 
 = \frac{1}{Z}
\sum_{i}e^{-\beta E_{i}}\langle i |n_{ {\bf{k}} }| i \rangle
 = \frac{1}{Z}
\sum_{i}e^{-\beta E_{i}}\sum_{ j_{0} }
 \langle i |n_{ {\bf{k}} }| j_{0} \rangle
\langle j_{0} | i \rangle
= \frac{1}{Z}\sum_{i}e^{-\beta E_{i}}\sum_{ j_{0} }
n^{j0}_{ {\bf{k}} }|\langle j_{0} | i \rangle|^{2}
\end{equation}
Let us introduce the inverse of the "matrix" $\{ n^{j0}_{ {\bf{k}} } \} $
\begin{equation}
\sum_{ j0 }n^{j0}_{ {\bf{k}} }(n^{-1})_{j0}^{ {\bf{k}}^{'} }
 = \delta_{ {\bf{k}}, {\bf{k}}^{'} }
\end{equation}
\begin{equation}
\sum_{ {\bf{k}} }n^{i0}_{ {\bf{k}} }(n^{-1})_{j0}^{ {\bf{k}} }
 = \delta_{i0,j0 }
\end{equation}
Therefore,
\begin{equation}
\sum_{ {\bf{k}} }
\langle n_{ {\bf{k}} } \rangle (n^{-1})_{j0}^{ {\bf{k}} }
= \frac{1}{Z}\sum_{i}e^{-\beta E_{i}}|\langle j_{0} | i \rangle|^{2}
\end{equation}
\begin{equation}
\langle n_{ {\bf{k}} }n_{ {\bf{k}}^{'} } \rangle
 =  \sum_{ j_{0} }
(\frac{1}{Z}
\sum_{i}e^{-\beta E_{i}}|\langle j_{0} | i \rangle|^{2})
n^{j0}_{ {\bf{k}} }n^{j0}_{ {\bf{k}}^{'} }
\end{equation}
Therefore,
\begin{equation}
\langle n_{ {\bf{k}} }n_{ {\bf{k}}^{'} } \rangle
 =  \sum_{ j_{0} }\sum_{ {\bf{p}} }
\langle n_{ {\bf{p}} } \rangle (n^{-1})_{j0}^{ {\bf{p}} }
n^{j0}_{ {\bf{k}} }n^{j0}_{ {\bf{k}}^{'} }
\end{equation}
Thus,
\begin{equation}
N( {\bf{k}},{\bf{k}}^{'} ) = 
\sum_{ j_{0} }\sum_{ {\bf{p}} }
\langle n_{ {\bf{p}} } \rangle (n^{-1})_{j0}^{ {\bf{p}} }
n^{j0}_{ {\bf{k}} }n^{j0}_{ {\bf{k}}^{'} }
- \langle n_{ {\bf{k}} } \rangle\langle n_{ {\bf{k}}^{'} } \rangle
\end{equation}
 The above formula is exclusively in terms of the (nonideal) 
 average $ \langle n_{ {\bf{p}} } \rangle $. The rest of the quantities
 are known except the "inverse" of $ n^{j0}_{ {\bf{k}}^{'} } $ may be
 hard to compute. This preprint together with our earlier one
 (cond-mat/9810043) presents a viable alternative (hopefully)
 to other approaches at computing single-particle properties of Fermi systems.

\end{document}